\documentclass[journal=jacsat,manuscript=article]{achemso}

\usepackage[version=3]{mhchem} 
\usepackage{physics}

\usepackage{subfigure}
\usepackage{subcaption}
\usepackage{subfig}

\usepackage{xcolor}

\usepackage{amsmath,amssymb}
\usepackage{pst-blur}
\usepackage{pstricks-add}
\definecolor{Blue}{rgb}{1.,0.75,0.8}
\usepackage{tabularx}
\usepackage{graphicx}

\author{Ariadni Boziki}
\affiliation{Department of Physics and Materials Science, University of Luxembourg, L-1511, Luxembourg City, Luxembourg}
\author{Frédéric Ngono Mebenga}
\affiliation{Janssen Research and Development, Pharmaceutical and material sciences (PM\&S), Janssen Pharmaceutical Companies of Johnson \& Johnson, 2340 Beerse, Belgium}
\author{Philippe Fernandes}
\affiliation{Janssen Research and Development, Pharmaceutical and material sciences (PM\&S), Janssen Pharmaceutical Companies of Johnson \& Johnson, 2340 Beerse, Belgium}
\author{Alexandre Tkatchenko}
\affiliation{Department of Physics and Materials Science, University of Luxembourg, L-1511, Luxembourg City, Luxembourg}
\email{alexandre.tkatchenko@uni.lu}

\title[An \textsf{achemso} demo]
  {A Journey with THeSeuSS: Automated Python Tool for Modeling IR and Raman Vibrational Spectra of Molecules and Solids}

\abbreviations{IR,NMR,UV}

\begin{document}



\begin{abstract}
Vibrational spectroscopy is an indispensable analytical tool that provides structural fingerprints for molecules, solids, and interfaces thereof. This study introduces THeSeuSS (THz Spectra Simulations Software) -- an automated computational platform that efficiently simulates IR and Raman spectra for both periodic and non-periodic systems. Utilizing DFT and DFTB, THeSeuSS offers robust capabilities for detailed vibrational spectra simulations. Our extensive evaluations and benchmarks demonstrate that THeSeuSS accurately reproduces both previously calculated and experimental spectra, enabling precise comparisons and interpretations of vibrational features across various test cases, including H$_{2}$O and glycine molecules in the gas phase, as well as solid ammonia and solid ibuprofen. Designed with a user-friendly interface and seamless integration with existing computational chemistry tools, THeSeuSS enhances the accessibility and applicability of advanced spectroscopic simulations, supporting research and development in chemical, pharmaceutical, and material sciences. Future updates aim to expand its methodological diversity by incorporating machine learning techniques to analyze larger and more complex systems, solidifying THeSeuSS's role as an essential tool in the computational chemist’s arsenal.
\end{abstract}

\section{Introduction}

In various industries including chemical, petrochemical, polymer, pharmaceutical, cosmetic, food, and agricultural sectors, there is a pressing demand to enhance product quality and optimize production processes. This has led to a notable resurgence in the application of vibrational spectroscopy.\cite{larkin2017infrared, schrader1995infrared} Recent years have witnessed groundbreaking advancements in coherent nonlinear vibrational spectroscopy techniques, including Infrared (IR), THz, IR-Raman, IR-vis, vis-IR, and THz-Raman techniques.\cite{doi:10.1021/acs.chemrev.9b00813, mukamel1995principles, hamm2011concepts, 10.1063/1.477756, 10.1063/1.4916522, 10.1063/1.4932983}. Both IR and Raman spectroscopy provide complementary insights into molecular vibrations: IR absorption is sensitive to anti-symmetric vibrations that induce changes in dipole moment, while Raman scattering is sensitive to symmetric vibrations that alter polarizability.\cite{larkin2017infrared, schrader1995infrared}. The resulting vibrational spectrum acts as a unique ``fingerprint'', enabling the characterization, quantification, and elucidation of intrinsic features of functional groups or entire compounds.\cite{doi:10.1080/05704928.2016.1230863, doi:10.1080/05704920701551530, doi:10.1021/acs.chemrev.5b00640, ho2008signatures}

Despite notable advancements, the comprehension of vibrational spectra remains in its early stages, necessitating further exploration to unlock the nature of spectral features and the underlying structural information. Specifically, deciphering the vibrational spectra of complex molecules, particularly those with numerous degrees of freedom, poses a challenge. The task of assigning peaks in the spectra to specific vibrational modes typically requires a combination of experience and chemical intuition. However, the complexity inherent in these spectra often renders such assignments impractical or even unattainable.\cite{https://doi.org/10.1002/jcc.10089} High-level quantum chemistry methods have demonstrated their reliability in accurately simulating vibrational spectra, even for large molecules.\cite{doi:10.1021/ac2001934, doi:10.1021/acs.jpca.6b05702, WANG201899, doi:10.1021/acs.molpharmaceut.2c00509, doi:10.1021/jp107597q, doi:10.1021/acs.jpca.9b00792, C6CP07388C, ruggiero2020invited, doi:10.1021/acs.jctc.0c00127, Azuri2018, PhysRevLett.113.055701, PhysRevLett.119.097404, https://doi.org/10.1002/wcms.1605, billes2015application, C1CP21830A, https://doi.org/10.1002/wcms.1480}

A variety of theoretical methods have been utilized to simulate vibrational spectra, ranging from highly accurate approaches like coupled-cluster CCSD(T) and quadratic configuration interaction to more computationally efficient techniques such as Density Functional Theory (DFT) and classical force fields.\cite{DARGELOS2020137746, doi:10.1021/jp952471r, https://doi.org/10.1002/qua.22707, Palafox+2018, doi:10.1021/ct100326h, https://doi.org/10.1002/qua.20469, CLARKSON2003413, SUNDARAGANESAN2007628, doi:10.1021/acs.jctc.0c00127, 10.1063/1.1670007, doi:10.1021/jp412827s} However, the computational cost associated with methods such as coupled-cluster makes them impractical for computing vibrational frequencies of realistically sized molecules, rendering such calculations intensive and often infeasible. Consequently, DFT has emerged as a widely adopted approach due to its comparatively lower computational demands, making it the method of choice for many theoretical vibrational spectroscopy studies.\cite{doi:10.1021/ac2001934, doi:10.1021/acs.jpca.6b05702, doi:10.1021/acs.molpharmaceut.2c00509, doi:10.1021/jp107597q, doi:10.1021/ct100326h, https://doi.org/10.1002/qua.20469} Yet, when investigating large systems like biomolecules or polymers, even DFT becomes computationally prohibitive. In such cases, alternative approaches such as the density-functional tight-binding (DFTB) method, machine learning models, and classical force fields provide viable solutions for calculating vibrational spectra in complex systems.\cite{10.1063/1.1775787, doi:10.1021/jp0742822, doi:10.1142/S0219633605001763, 10.1063/1.2806992, MALOLEPSZA2005237, doi:10.1021/acs.jpca.1c10417, doi:10.1021/jacs.0c06530, doi:10.1021/acs.jctc.8b00524, C7SC02267K, Raimbault_2019,  doi:10.1021/acs.jctc.0c00127, 10.1063/1.1670007, doi:10.1021/jp412827s}

Many electronic structure and classical molecular dynamics codes offer the capability to calculate vibrational frequencies, with some also capable of computing IR intensities and Raman activities. However, these capabilities are less prevalent in codes designed for periodic systems. In such cases, many packages rely on external tools like PHONOPY for phonon frequency calculations, rather than incorporating built-in functions or libraries.\cite{doi:10.7566/JPSJ.92.012001, Togo_2023} Examples of codes adept at simulating vibrational/phonon frequencies and/or IR intensities and Raman activities, include FHI-aims,\cite{BLUM20092175, HAVU20098367, Marek_2014, YU2018267, YU2020107459} Quantum ESPRESSO,\cite{Giannozzi_2009, Giannozzi_2017} Gaussian,\cite{g16} GAMESS,\cite{GAMESS} ORCA,\cite{10.1063/5.0004608} TURBOMOLE,\cite{https://doi.org/10.1002/wcms.1162} CP2K,\cite{10.1063/5.0007045} CASTEP,\cite{ClarkSegallPickardHasnipProbertRefsonPayne+2005+567+570} DFTB+,\cite{10.1063/1.5143190} LAMMPS,\cite{THOMPSON2022108171} TRAVIS,\cite{C3CP44302G, C4CP05272B} QERaman,\cite{HUNG2024108967} MTASpec\cite{KHIRE2022108175} and others. Moreover, there is a shortage of codes equipped to concurrently compute IR and Raman spectra at various levels of theory, such as DFT and DFTB. Typically, these codes specialize in either quantum mechanical or semiclassical/fully classical spectrum calculations.

Here, we introduce the THeSeuSS (THz Spectra Simulations Software), a sophisticated Python tool designed for simulating IR and Raman spectra utilizing a static method (diagonalization of the dynamical matrix) at both DFT and DFTB levels of theory. Our package offers a fully automated platform, streamlining the spectra calculation process. Users simply provide the experimental crystal structure and input parameters to THeSeuSS, enabling computation of spectra for both periodic and non-periodic systems. THeSeuSS interfaces with FHI-aims and DFTB+ codes to handle electronic structure calculations. In contrast to the built-in functionalities of FHI-aims for IR and Raman spectra calculations, our implementation significantly reduces computational time. To optimize calculations of periodic systems, our code interfaces with PHONOPY, leveraging its incorporation of symmetry operations. By selectively displacing coordinates during the finite displacement method, we minimize computational overhead, requiring only calculations for force, polarizability, and Cartesian polarization of irreducible representations. Furthermore, unlike DFTB+, our code extends its capabilities to compute IR intensity and Raman activity for periodic systems. Moreover, building upon the aforementioned optimization strategies, the single-point calculations linked to the finite displacement method are fully parallelized, thereby further improving computational efficiency.

\section{Methodology}

\subsection{IR and Raman spectroscopy}

IR and Raman spectroscopy both involve the interaction between radiation and molecular vibrations. However, they vary in their mechanisms for transferring photon energy to the molecule, which alters its vibrational state. IR spectroscopy relies on transitions between molecular vibrational levels induced by the absorption of mid-IR radiation, a process governed by resonance conditions that involve the electric dipole-mediated transition between energy levels. For a vibrational mode to be deemed "IR active", it must induce changes in the dipole moment without necessarily possessing a permanent dipole.\cite{stuart2004infrared, stuart2000infrared}

Raman spectroscopy involves a two-photon inelastic light scattering event, where the incident photon carries significantly more energy than the vibrational quantum energy. Part of this energy is transferred to the molecular vibration, causing the remaining energy to be scattered as a photon with reduced frequency. This interaction, characterized by the Raman polarizability of the molecule, occurs under off-resonance conditions. Within a Raman spectrum, peaks represent the intensity and wavelength of Raman-scattered light, each corresponding to distinct molecular bond vibrations spanning individual and group bonds to lattice modes. Notably, the intensity of Raman spectrum peaks is directly linked to both concentration and changes in polarizability.\cite{smith2019modern, keresztury2006r} 

\subsection{Calculation of vibrational/phonon frequencies}

Theoretical simulations employ two primary methods to determine vibrational spectra frequencies: the static and dynamic approaches.\cite{https://doi.org/10.1002/wcms.1605} The static approach, where nuclear positions remain fixed, is commonly employed in various spectroscopic techniques. However, for simulating vibrational spectra of nonrigid and large molecules, exploring the surrounding region of the equilibrium structure and averaging properties of interest from multiple calculations can prove advantageous. Conversely, under certain mechanisms and conditions such as finite temperature, the dynamic approach is considered superior. In the dynamic approach, nuclear positions evolve according to classical or quantum dynamics, necessitating either first-principles or classical molecular dynamics simulations. While first-principles molecular dynamics simulations are computationally expensive, especially for large systems, classical molecular dynamics offer an alternative. However, empirical force fields typically apply to specific compounds, such as proteins or polymers, rather than universal models applicable to all systems under study. Given our aim to accommodate systems of varying chemical environment, sizes and periodicity, we opt for the static approach to calculate vibrational/phonon frequencies.

In the static approach, the nuclear positions are initially optimized at 0 K, followed by displacements of coordinates using either the finite displacement method, or Density Functional Perturbation Theory (DFPT).\cite{heine1970solid, PhysRevLett.58.1861, PhysRevB.39.13120, PhysRevB.43.7231, PhysRevB.54.16470, PhysRevB.55.10355, RevModPhys.73.515} From a computational cost perspective, DFPT holds an advantage, particularly for periodic systems, as it enables the exact calculation of phonons at any wave vector. However, the finite displacement method offers a practical advantage. Many popular first-principles codes include implementations for atomic force calculations using various exchange-correlation functionals, pseudo-potential methods, and beyond-DFT approaches, making it readily accessible. In contrast, DFPT implementations may not be as widely available. To ensure user flexibility and accommodate various calculation preferences, the THeSeuSS package employs the finite difference method for calculating vibrational/phonon frequencies.

In perfectly ordered periodic crystals, both IR and Raman spectroscopy reveal distinct peaks corresponding to vibrational modes located at the zone center, a phenomenon attributed to crystal momentum conservation.\cite{10.1063/5.0017496} Phonon calculations carried out at the $\Gamma$-point serve as a suitable approach for deciphering the IR and Raman spectra, based on the premise of momentum conservation. This method considers the negligible momentum of visible and IR light in comparison to the size of the Brillouin zone.\cite{10.1063/5.0060718}

\subsubsection{Finite difference method}

The potential energy of a molecule or crystal, denoted as $\mathcal{V(\textbf{R})}$, where $\textbf{R}$ represents the atomic positions, can be expanded into a Taylor series. In the harmonic approximation, we approximate the total energy around the equilibrium positions by neglecting terms higher than second order:
\begin{equation}
    \mathcal{V}_{Harmonic} = \mathcal{V}\big (\{\textbf{R}^\xi_0\}_{\xi\in\text{atoms}}\big ) + \frac{1}{2} \sum_{\alpha,\beta}\sum_{i,j}\frac{\partial^2 \mathcal{V}(r_0,\cdots,r_{3|\text{atoms}|})}{\partial R_{i}^{\alpha} \partial R_{j}^{\beta}} \biggr\rvert_{\{\textbf{R}^{\xi}_0\}_{\xi \in \text{atoms}}} \Delta R_{i}^{\alpha} \Delta R_{j}^{\beta}
    \label{eq.18}
\end{equation}
where the atoms are indexed with $\alpha,\beta$ and $i,j=x,y,z$ are cartesian indices. The linear term is eliminated as the system experiences no forces while in equilibrium.

While first derivatives are relatively straightforward to compute, the same cannot be said for second derivatives. Equation \ref{eq.18} involves the second derivative of energy with respect to atomic displacements from equilibrium, known as the matrix of force constants or the Hessian. The Hessian elements, $\mathcal{H}_{i,j}^{\alpha,\beta} = \frac{\partial^2 \mathcal{V}}{\partial R_{i}^{\alpha} \partial R_{j}^{\beta}}$, can be assessed by numerically computing the second-order derivative using finite differences.
\begin{equation}
    \mathcal{H}_{i,j}^{\alpha,\beta} = \frac{\partial^2 \mathcal{V}}{\partial R_{i}^{\alpha} \partial R_{j}^{\beta}}\biggr\rvert_{{\{\textbf{R}^{\xi}_0\}_{\xi \in \text{atoms}}}} = - \frac{\partial} {\partial R_{i}^{\alpha}} F_{j}^{\beta} \biggr\rvert_{{\{\textbf{R}^{\xi}_0\}_{\xi \in \text{atoms}}}} = - \lim_{\epsilon\to0} \sum_{\hat{\iota}}\frac{F_{j}^{\beta}(\textbf{R}^{\alpha}_{0} + \epsilon \hat{\iota})}{\epsilon},
\end{equation}
where $F_{j}^{\beta}$ represents the force applied to the coordinate $j$ of atom $\beta$ when coordinate $i$ of atom $\alpha$ is displaced and $\hat{\iota}$ stands for the unit axial directional vectors.

To compute the second derivative, this process is repeated for all 3N Cartesian coordinates. Each atomic coordinate undergoes incremental adjustments: first, it is increased by a small amount ($+\frac{1}{2}\delta$), where $\delta$ is the displacement, and the gradients are calculated. Then, the coordinate is decreased ($-\frac{1}{2}\delta$), and the gradients are recalculated. The second derivative is then determined using the two-point central difference formula.
\begin{equation}
    \mathcal{H}_{i,j}^{\alpha,\beta} = \sum_{\hat{\iota}}\frac{\textbf{F}^{\alpha}(\textbf{R}^{1}_{0},\cdots,\textbf{R}^{\beta}_{0}+\hat{\iota}\frac{\delta}{2},\cdots)-\textbf{F}^{\alpha}(\textbf{R}^{1}_{0},\cdots,\textbf{R}^{\beta}_{0}-\hat{\iota}\frac{\delta}{2},\cdots)}{\delta}
    := \frac{1}{\delta}\sum_{\hat{\iota}}\Delta_{\beta \hat{\iota}} \mathbf{F}^\alpha
\end{equation}
Because the Hessian is symmetric, $\mathcal{H}_{i,j} = \mathcal{H}_{j,i}$, random errors arising from gradient calculations can be mitigated (by a factor of $(1/2)^{(1/2)}$) by redefining the Hessian as follows:
\begin{equation}
    \mathcal{H}^{\alpha,\beta}_{i,j} = \frac{1}{2\delta}\sum_{\hat{\iota}}\left(\Delta_{\alpha \hat{\iota}} \mathbf{F}^\beta + \Delta_{\beta \hat{\iota}} \mathbf{F}^\alpha\right)
    \label{eq.20}
\end{equation}
This matrix represents the force constants for the system. Prior to calculating the vibrational frequencies, it must be mass-weighted:
\begin{equation}
    \mathcal{H}_{i,j}^{\alpha,\beta} = \frac{\mathcal{H}_{i,j}^{\alpha,\beta}}{\sqrt{M_{i}^{\alpha} M_{j}^{\beta}}}.
\end{equation}
By diagonalizing the mass-weighted Hessian matrix, the eigenvectors $\{Q_l\}$ of it represent the vibrational normal modes, while the corresponding eigenvalues provide the vibrational frequencies $\{\omega_l\}$.

However, an added complexity arises when dealing with periodic boundary conditions, as we must also consider the atoms within the unit cell along with their periodic images. In the finite-displacement supercell approach, first-principles calculations serve as the foundation for obtaining atomic forces within a supercell crystal structure model. Force constants are derived from multiple supercells, each subjected to different sets of displacements. Phonons are computed exactly (i.e., without interpolation) from the force constants of the supercell at wave vectors commensurate with its shape relative to the primitive cell. For other wave vectors, phonons are obtained through interpolation. Typically, using a supercell size containing a few hundred atoms yields reasonable phonon results through interpolation, although the required accuracy may vary depending on the specific calculation objectives.\cite{doi:10.7566/JPSJ.92.012001} In our approach phonon frequencies are calculated utilizing PHONOPY package. A detailed mathematical description on how to calculate the phonon frequencies of periodic systems is given by \textit{Togo}. \cite{doi:10.7566/JPSJ.92.012001}

\subsection{IR intensity}

The standard IR spectrometer employs a broad-band source that simultaneously emits all IR frequencies of interest. These frequencies cover the near-IR region (14,000-4000 cm$^{-1}$), the mid-IR region (4000-400 cm$^{-1}$), and the far-IR region (400-10 cm$^{-1}$).\cite{larkin2018ir} As previously discussed, a well-known selection rule in IR spectroscopy dictates that a change in dipole moment during vibration renders the mode IR active.\cite{PhysRevB.54.7830, https://doi.org/10.1002/jcc.10089} More specifically, IR absorption occurs via an electric dipole operator-mediated transition, wherein the change in dipole moment relative to a change in vibrational amplitude is greater than zero.

The IR intensity, denoted as $I_{i}$, for the $i$th vibrational normal mode within the double-harmonic approximation, can be expressed as follows:
\begin{equation}
    I_i \propto \Bigg (\frac{\partial \boldsymbol{\mu}}{\partial Q_{i}} \Bigg )^{2}.
    \label{eq.11}
\end{equation}
Here, we omit proportionality factors and quantum corrections, $Q_{i}$ represents the $i$th normal coordinate and $\boldsymbol{\mu}$ denotes the system's dipole moment.\cite{https://doi.org/10.1002/wcms.1605} The definition of $\boldsymbol{\mu}$ is typically applicable only to isolated molecules, where periodic boundary conditions are absent. However, in periodic systems, the Berry phase approach to polarization and maximally localized Wannier functions (MLWFs) have proven successful in addressing this limitation.\cite{PhysRevB.47.1651, RevModPhys.66.899, PhysRevB.56.12847, 10.1063/1.479638, PhysRevLett.82.3308}

To determine the IR intensity (as defined in Eq. \ref{eq.11}), our methodology involves several steps. Initially, we compute the dipole moment for both FHI-aims and DFTB+ in non-periodic systems. For periodic systems, in case of FHI-aims, we calculate the Cartesian polarization utilizing the Berry phase approach to polarization and MLWFs, while in the case of DFTB+ we calculate the dipole moment for periodic systems. Following this, we determine the derivative through the application of finite differences.\cite{PhysRevMaterials.3.053605, PhysRevB.55.10337}

\subsection{Raman activity}

Raman spectroscopy operates on the principle of inelastic light scattering, spanning the near-infrared, visible, and near-ultraviolet regions of the electromagnetic spectrum. As previously mentioned, for Raman bands to be detected, molecular vibrations must induce a change in polarizability. Thus, a fundamental component crucial for simulating Raman intensity is the electronic static polarizability tensor, represented as $\Tilde{\alpha}$.

In Raman spectroscopy, two types of Raman scattering are observed: Stokes and anti-Stokes. Stokes Raman scattering occurs when molecules initially in the ground vibrational state undergo scattering, while anti-Stokes Raman scattering occurs when molecules initially in a vibrational excited state undergo scattering.\cite{larkin2018ir} In most experimental setups, a plane-polarized laser beam serves as the incident beam, with the incident beam's direction, polarization orientation, and the observation direction all mutually perpendicular. Under these conditions, the first-order differential Raman cross-section, specifically for the Stokes component of the $i$th eigenmode far from resonance, within the double-harmonic approximation, is as follows:\cite{PhysRevB.54.7830}
\begin{equation}
    \frac{\partial\sigma_i}{\partial\Omega} = \frac{(2\pi\nu_s)^4}{c^4} \frac{h(n_{i}^{b} + 1)}{8\pi^2\nu_{i}} \frac{I^{\text{Ram}}}{45}
    \label{eq.14}
\end{equation}
where $\nu_{s}$ and $\nu_{i}$ are the wavenumbers of the scattered light and the $i$th normal coordinate, respectively, $n_{i}^{b}$ is the Bose-Einstein statistical factor, expressed as:
 \begin{equation}
     n_{i}^{b} = \Bigg [ \exp{\frac{h\nu_{i}}{kT}} -1 \Bigg ]^{-1}.
     \label{eq.13}
 \end{equation}
and $I^{\text{Ram}}$ is the Raman-scattering activity:\cite{PhysRevB.54.7830}
\begin{equation}
    I^{\text{Ram}} = 45 \bigg ( \frac{\partial\alpha}{\partial Q} \bigg )^2 + 7 \bigg ( \frac{\partial\beta}{\partial Q} \bigg )^2 = 45\alpha'^2 + 7\beta'^2.
    \label{eq.15}
\end{equation}
In Eq. \ref{eq.15}, $\alpha'$ is the mean polarizability derivative and $\beta'^2$ is the anisotropy of the polarizability tensor derivative. The expressions for $\alpha'$ and $\beta'^2$ are given by:
\begin{equation}
    \alpha' = \frac{1}{3}(\Tilde{\alpha}'_{xx} + \Tilde{\alpha}'_{yy} + \Tilde{\alpha}'_{zz}),
    \label{eq.16}
\end{equation}
\begin{equation}
    \beta'^{2} = \frac{1}{2}[(\Tilde{\alpha}'_{xx}-\Tilde{\alpha}'_{yy})^{2} + (\Tilde{\alpha}'_{xx}-\Tilde{\alpha}'_{zz})^{2} + (\Tilde{\alpha}'_{yy}-\Tilde{\alpha}'_{zz})^{2}] + 6(\Tilde{\alpha}'^{2}_{xy} + \Tilde{\alpha}'^{2}_{xz} + \Tilde{\alpha}'^{2}_{yz}).
    \label{eq.17}
\end{equation}
From a computational standpoint, the IR activity serves as the basis for our calculations in the Raman spectrum.

To determine the Raman activity (as defined in Eq. \ref{eq.15}), we undertake several steps. Initially, we compute the polarizability matrix for both FHI-aims and DFTB+ in both periodic and non-periodic systems. For FHI-aims, we employ DFPT, while for DFTB+, polarizability is evaluated using coupled-perturbed linear response (CPLR).\cite{10.1063/5.0137122} Subsequently, akin to the procedure for IR intensity determination, we derive the necessary derivative using finite differences.\cite{PhysRevMaterials.3.053605, PhysRevB.55.10337}

\section{Features of the code}

THeSeuSS is a fully automated tool designed to streamline the calculation of IR and Raman spectra for a wide range of systems, from small molecules to large molecules and crystals. It consists of the following components:

\begin{enumerate}
\item \textbf{Conversion of Experimental Geometry Input}:
Converts .cif format files containing experimental structures into either FHI-aims or DFTB+ geometry input files, (only for periodic systems).
\item \textbf{Space Group Determination of the Experimental Structure}: Determines the space group of the experimental crystal structure, (only for periodic systems). This determination is facilitated by the \verb|Spglib| library, a software for crystal symmetry search.\cite{togo2018texttt}
\item \textbf{Geometry Optimization}: Optimizes the geometry for both periodic and non-periodic systems as well as the cell for periodic systems. Depending on the chosen method, (whether DFT or DFTB), users have access to a wide range of functionals, dispersion method models, and simulation parameters provided by FHI-aims for DFT calculations. Similarly, for DFTB calculations, users can utilize various models for dispersion interactions and simulation parameters provided by DFTB+ software.
\item \textbf{Space Group Determination of the Optimized Structure}: Identifies the space group of the optimized crystal structure, (only for periodic systems).
\item \textbf{Vibrational/Phonon Frequency Calculations}:
Utilizes the finite difference method to compute vibrational or phonon frequencies. For non-periodic systems, THeSeuSS includes built-in functions, while for periodic systems, the PHONOPY code is employed.\cite{Togo_2023, doi:10.7566/JPSJ.92.012001} This part, along with the following section, is the most time-consuming. To address this, we have parallelized these components. In essence, following the finite difference method described in the Methodology section, each atomic coordinate undergoes incremental adjustments, both increased and decreased by the same small amount. Subsequently, a single-point calculation is performed for each case to calculate the gradient using Equation \ref{eq.20}. This process is repeated for all coordinates of all atoms in the system under study, particularly for the non-periodic case. To accelerate the frequencies computation, we utilize the \verb|ThreadPoolExecutor| from Python's \verb|concurrent.futures| module, enabling parallel execution of the single point calculations. In the case of periodic systems, leveraging PHONOPY introduces an additional efficiency feature. PHONOPY inherently accounts for the symmetry of the crystal structure, thereby reducing the necessary single-point calculations to only those atomic coordinates that comprise the irreducible representation. This significantly reduces the number of calculations required for periodic systems.
\item \textbf{Property Calculations}: For both periodic and non-periodic systems, while the FHI-aims code is used, DFPT calculations are performed to determine the polarizability matrix. When employing the DFTB+ code, we instead conduct CPLR calculations for the same purpose. Additionally, for periodic systems analyzed with FHI-aims, cartesian polarization is calculated using the Berry phase method and MLWFs. In both coding environments, the dipole moment is determined for non-periodic systems. Moreover, when using DFTB+, the dipole moment for periodic systems is also assessed. We employ the same parallelization scheme described earlier (part 5). For periodic systems, the calculations of these properties are performed for coordinates comprising the irreducible representation, as defined by the symmetry operations applied by PHONOPY. Utilizing the \verb|Spglib| library,\cite{togo2018texttt} we match atoms to their counterparts within the irreducible representation, enabling the construction of the full matrix for subsequent calculations of IR intensity and Raman activity.
\item \textbf{IR Intensity and Raman Activity}: Computes the IR intensity from the dipole moment or the cartesian polarization and the Raman activity from the polarizability matrix.
\item \textbf{Spectra Calculation}:
Utilizes the acquired vibrational/phonon frequencies, IR intensity, and Raman activity to generate the system's IR and Raman spectra. In addition to the raw spectral data, THeSeuSS offers internal built-in functions for spectrum plotting, allowing users to apply either Gaussian or Lorentzian broadening based on their preference.
\end{enumerate}

The code can execute all of its components simultaneously or individually, depending on user preference. This choice is customizable within the input file. Comprehensive details regarding the input file and usage examples can be found in the THeSeuSS documentation.\cite{THeSeuS_documentation} Beyond its utility as a tool for simulating IR and Raman spectra, THeSeuSS serves as an automation tool for structure optimization. Users can input experimental structures, and with its built-in functions, THeSeuSS can convert .cif format files into suitable inputs for FHI-aims or DFTB+. Additionally, it can serve solely as a spectrum generation tool if users have already computed frequencies, IR intensities, and Raman activities using alternative methods. Furthermore, THeSeuSS is capable of calculating THz spectra, where only the frequencies of the spectrum are computed. In Scheme \ref{sch.1}, we provide a concise overview of the algorithmic protocol employed to generate the IR and Raman spectra.

\begin{scheme}[H]
  \centering
  \includegraphics[width=185mm]{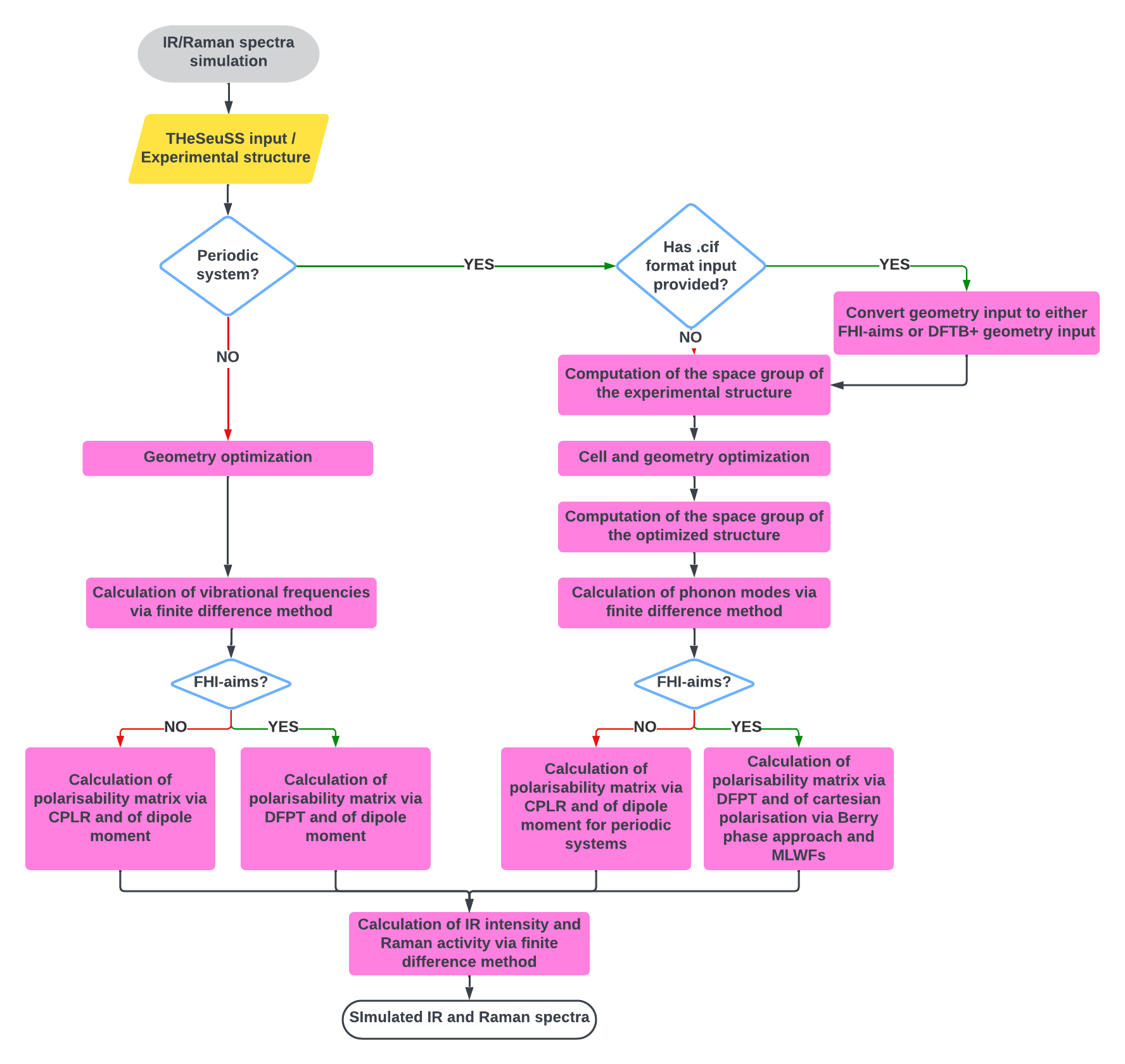}
  \caption{Algorithmic Protocol for Calculating IR and Raman Spectra: This entire process has been automated within our developed package THeSeuSS. The procedure begins with the optimization of the experimental crystal structure. Following this, we employ the finite difference method to compute vibrational/phonon frequencies. For FHI-aims, DFPT is employed to calculate the polarizability matrix for each displaced structure, while for DFTB+, we use the CPLR. Additionally, in the case of periodic systems using FHI-aims, the Berry phase method and MLWFs are applied to calculate the cartesian polarization. Once modes are classified as IR or Raman active, we calculate the IR intensity using the dipole moment/cartesian polarization and the Raman activity using the polarizability matrix. These calculated frequencies, alongside the IR intensity and Raman activity, are then utilized to generate the simulated spectra.}
  \label{sch.1}
\end{scheme}

\section{Evaluating and benchmarking THeSeuSS}

To evaluate the capability of the code in generating vibrational spectra, we selected both non-periodic and periodic systems for testing and benchmarking. For the non-periodic systems, we selected an H$_{2}$O molecule and a glycine molecule, both in the gas phase. For the periodic systems, we chose ammonia molecular crystal from the X23 database,\cite{10.1063/1.4738961, reilly2013understanding} and ibuprofen [2-(4-isobutylphenyl) propionic acid] molecular crystal, known for its nonsteroidal anti-inflammatory, analgesic, and antipyretic properties.

For the tests and benchmarks associated with the FHI-aims code in periodic systems, we have chosen the PBE functional,\cite{PhysRevLett.77.3865} supplemented by the nonlocal many-body dispersion (MBD-NL) method.\cite{PhysRevLett.124.146401} This selection is backed by extensive benchmarks established by the Crystallographic Data Centre (CCDC) through blind tests that assess computational methods capable of precisely describing the intricate nature of molecular crystals.\cite{reilly2016report, day2009significant, bardwell2011towards, doi:10.1126/sciadv.aau3338} Although these benchmarks highlight the hybrid PBE0 functional,\cite{adamo1999toward} along with the MBD approach,\cite{PhysRevLett.108.236402, distasio2012collective, ambrosetti2014long, distasio2014many} we opted for the PBE functional due to its lower computational demands for DFPT calculations of the polarizability matrix, which are generally four times more resource-intensive than a single force evaluation.\cite{PhysRevMaterials.3.053605, Raimbault_2019} For non-periodic systems, we utilize several functionals (PBE, PBE0, B3LYP)\cite{PhysRevB.37.785, 10.1063/1.464913} both with and without the MBD-NL method, with some results reported in the Supporting Information. For the DFTB+ code, we employed the MBD method and performed vibrational spectra simulations at the third-order Density Functional Tight Binding (DFTB3+MBD) level.

\begin{figure}[h!]%
\centering
\subfigure[IR spectra of H$_{2}$O molecule in gas phase.]{
\includegraphics[height=2.5in]{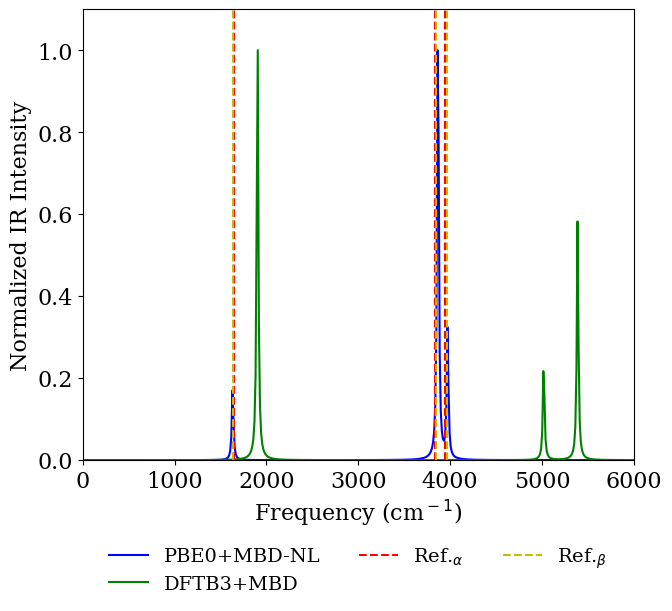}}%
\qquad
\subfigure[Raman spectra of H$_{2}$O molecule in gas phase.]{%
\includegraphics[height=2.5in]{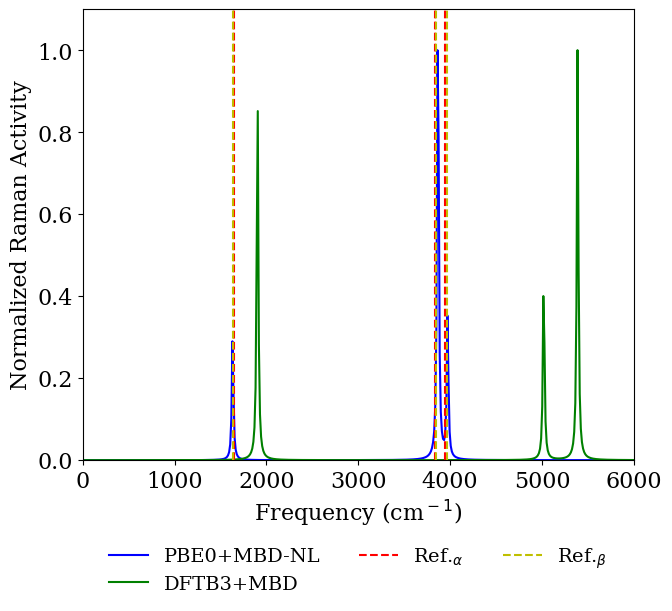}}%
\caption{Simulated IR and Raman spectra of a H$_{2}$O molecule in the gas phase, calculated at both the PBE0+MBD-NL and DFTB3+MBD levels, compared with reference data.\cite{doi:10.1021/ct500860v} Abbreviations: Ref.$_{\alpha}$: reference data obtained with CCSD(T) and ha5Z basis set, Ref.$_{\beta}$: reference data obtained with MP2 and ha5Z basis set.}
\label{fig:5}%
\end{figure}

In Figure \ref{fig:5}, we present the simulated IR and Raman spectra of a H$_{2}$O molecule in the gas phase. To assess the accuracy of our approach in capturing vibrational spectra, we compare our results with data from previous calculations.\cite{doi:10.1021/ct500860v} This choice is driven by the nature of available experimental data, which predominantly feature H$_{2}$O clusters in liquid, gas, or solid states, rather than isolated molecules. In these states, hydrogen-bonding networks break the symmetry of the molecular structure, leading to qualitatively different vibrations compared to isolated molecules. This phenomenon is also evident in several studies that report computed vibrational frequencies of small water clusters at the CCSD(T) level.\cite{doi:10.1021/ct500860v, doi:10.1021/acs.jctc.5b00225, 10.1063/1.4829463, 10.1063/1.4820448, 10.1063/1.3554905, 10.1063/1.3196178} These studies primarily focus on calculating vibrational frequencies and other properties for small water clusters, (H${_2}$O)${_n}$, where $n=2–6$. They demonstrate that as the cluster size increases from dimer to trimer and beyond, the vibrational frequencies differ for each distinct water cluster. As detailed in Table \ref{table:1}, in the gas phase, the H$_{2}$O molecule displays two types of stretching vibrations (symmetric and antisymmetric) and a bending vibration.\cite{10.1063/1.5037113, ZOBOV1996381, FALK198443, doi:10.1021/ct500860v} To quantitatively compare our simulated spectra with reference data, we use the mean absolute percentage error (MAPE), calculated as follows:
\begin{equation}
    \text{MAPE} = \frac{100}{N} \times \sum_{i=1}^N \abs{\frac{y_i-\hat{y_i}}{y_i}}.
\end{equation}
Additionally, we assess the frequency shifts as percentages, defined by:
\begin{equation}
    \text{Shift Percentage} = 100 \times \abs{\frac{y_i-\hat{y_i}}{y_i}}
\end{equation}
In these equations, $y_{i}$ represents the reference data, and $\hat{y_{i}}$ denotes our simulated results.

Table \ref{table:1} demonstrates that the simulated spectra at the PBE0+MBD-NL level show excellent agreement with the reference data, particularly in accurately describing vibrational frequencies. We also calculated vibrational frequencies using PBE, PBE+MBD-NL, PBE0 and B3LYP, as detailed in Table S1 of the Supporting Information. As expected, the results from hybrid functionals (PBE0, B3LYP) outperform those obtained with the PBE functional. The use of MBD-NL did not impact the results, as this method primarily addresses long-range dispersion forces. However, it is important to note that hybrid functionals (PBE0 and B3LYP) are not implemented in the DFPT framework of FHI-aims. Consequently, when evaluating THeSeuSS performance with a hybrid functional, the polarizability matrix and in turn Raman activity are computed at the corresponding PBE or PBE+MBD-NL levels.

\begin{table}
\centering  
{\scriptsize
\begin{tabular}{|c c c c |}
    \hline
    PBE0+MBD-NL & DFTB3+MBD &  Ref.$_{\alpha}$ & Ref.$_{\beta}$ \\  
     \hline
    \multicolumn{4}{|c|}{Frequencies (cm$^{-1}$)} \\ 
    \hline
     \begin{tabular}{@{}c@{}}  1633 \\ 3868 \\ 3973 \end{tabular} & \begin{tabular}{@{}c@{}} 1903 \\ 5020 \\ 5389 \end{tabular} & \begin{tabular}{@{}c@{}} 1650 \\ 3835 \\ 3945 \end{tabular} & \begin{tabular}{@{}c@{}} 1632 \\ 3843 \\ 3970 \end{tabular} \\ 
    \hline
    \hline
        \multicolumn{4}{|c|}{Frequency shifts (\%) with respect to Ref.$_{\alpha}$} \\ 
    \hline
       \begin{tabular}{@{}c@{}} 1.03 \\ 0.86 \\ 0.71 \end{tabular} & \begin{tabular}{@{}c@{}} 15.33 \\ 30.90 \\ 36.60 \end{tabular} & \begin{tabular}{@{}c@{}} - \\ - \\ - \end{tabular} & \begin{tabular}{@{}c@{}} - \\ - \\ -  \end{tabular} \\ 
      \hline
    \multicolumn{4}{|c|}{MAPE (\%) with respect to Ref.$_{\alpha}$} \\
    \hline
       0.87 & 27.61 & - & - \\ 
      \hline
      \hline
        \multicolumn{4}{|c|}{Frequency shifts (\%) with respect to Ref.$_{\beta}$} \\ 
    \hline
       \begin{tabular}{@{}c@{}} 0.06 \\ 0.65 \\ 0.08 \end{tabular} & \begin{tabular}{@{}c@{}} 16.61 \\ 30.63 \\ 35.74 \end{tabular} & \begin{tabular}{@{}c@{}} - \\ - \\ - \end{tabular} & \begin{tabular}{@{}c@{}} - \\ - \\ -  \end{tabular} \\ 
      \hline
          \multicolumn{4}{|c|}{MAPE (\%) with respect to Ref.$_{\beta}$} \\
    \hline
      0.26  & 27.66 & - & - \\ 
      \hline
\end{tabular}
}
\caption{Performance benchmark of THeSeuSS for the H$_{2}$O molecule in the gas phase, quantified through MAPE and frequency shifts expressed as percentages. This table displays frequencies calculated with THeSeuSS alongside those from previously established calculations.\cite{doi:10.1021/ct500860v} Frequency shifts and MAPE are computed in comparison to these reference data. Abbreviations: Ref.$_{\alpha}$: reference data obtained with CCSD(T) and ha5Z basis set, Ref.$_{\beta}$: reference data obtained with MP2 and ha5Z basis set.}
\label{table:1} 
\end{table}

Conversely, the DFTB3+MBD method demonstrates more substantial deviations from the reference data, a result that aligns with expectations given the limitations of the DFTB approach arising by the semi-empirical nature of the method.\cite{https://doi.org/10.1002/wcms.1156, 10.1063/1.2778428, doi:10.1021/jp503372v} The deviations observed in our simulations using THeSeuSS should not be interpreted as deficiencies in our implementation. Indeed, several studies have been conducted to enhance the accuracy of DFTB parametrization for water.\cite{doi:10.1021/acs.jctc.9b00816, https://doi.org/10.1002/jcc.23677, 10.1063/5.0132903, doi:10.1021/acs.jpcb.3c03479} However, exploring these alternatives is beyond the scope of this paper; thus, we have focused solely on the DFTB3+MBD method without benchmarking different DFTB parametrizations. Despite these challenges, we will present all data related to DFTB to ensure a comprehensive analysis of our implementation.

Furthermore, we address the challenges of analyzing the vibrational spectra of the glycine molecule in the gas phase, primarily hindered by its low thermal stability. Achieving a sufficient gas-phase concentration of glycine to record its vibrational spectra is not feasible, which significantly impedes these studies. Researchers have often resorted to alternative methods, such as combining IR spectroscopy with matrix isolation techniques.\cite{doi:10.1021/jp973397a, KUMAR20052741, GRENIE1972240, SUZUKI19631195}

Quantum mechanical calculations reveal three stable conformers of glycine, referred to as conformers I, II, and III.\cite{doi:10.1021/jp973397a} If all three conformers are present in a sample, the resulting experimental vibrational spectrum would likely be a superposition of their individual spectra, adding a layer of complexity to accurate measurements. For our benchmarks, we focused on conformer I. This conformer's molecular point group, C$_{\text{s}}$, allows all 24 of its vibrational modes to be active in both IR and Raman spectroscopy.

\begin{figure}[h!]%
\centering
\subfigure[IR spectra of glycine molecule in gas phase.]{
\includegraphics[height=4in]{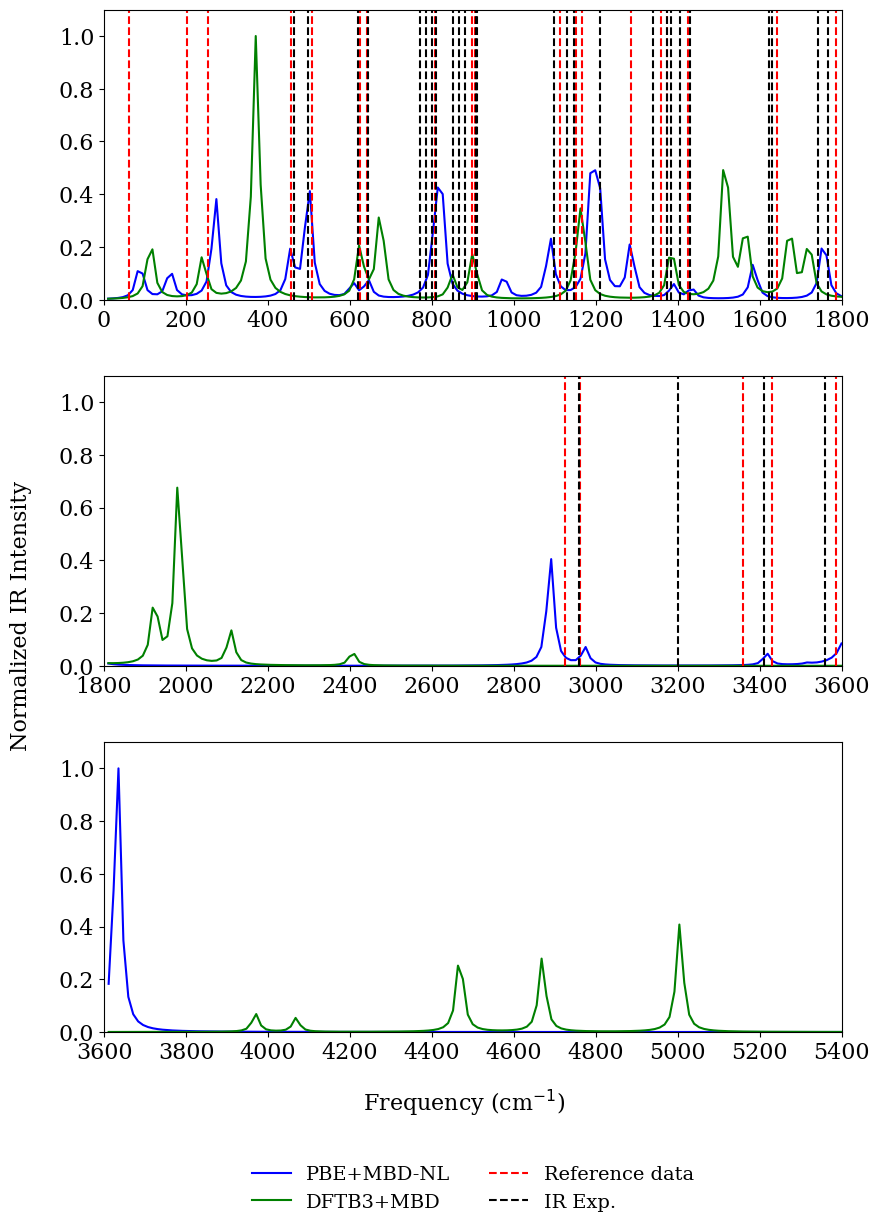}}%
\qquad
\subfigure[Raman spectra of glycine molecule in gas phase.]{
\includegraphics[height=4in]{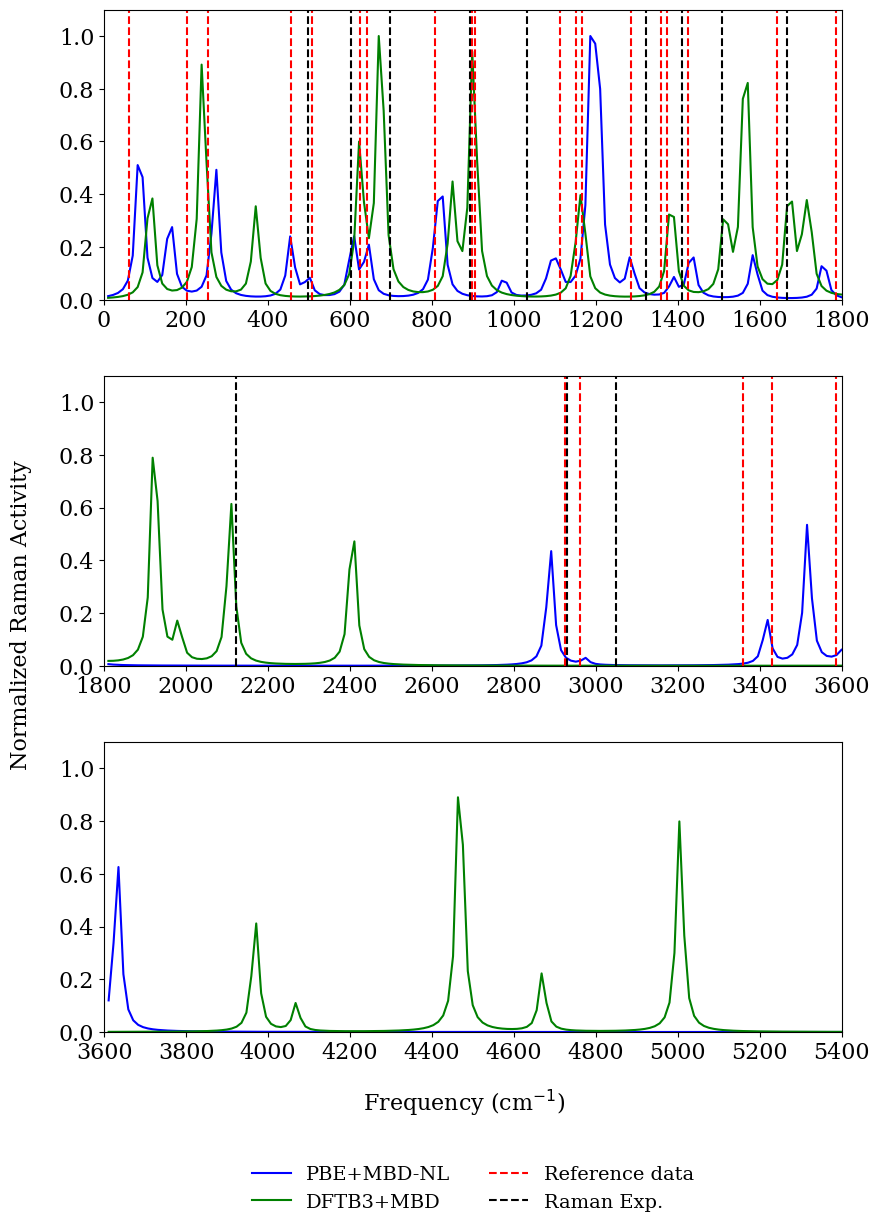}}%
\caption{Vibrational spectra of a glycine molecule in the gas phase: (a) IR spectra calculated using both PBE+MBD-NL and DFTB3+MBD levels, compared with previously established calculations and experimental IR frequencies.\cite{doi:10.1021/jp973397a} (b) Raman spectra calculated at both the PBE+MBD-NL and DFTB3+MBD levels, compared with previously established calculations and experimental Raman frequencies.\cite{doi:10.1021/jp973397a, KUMAR20052741} Abbreviations: IR Exp.: experimental IR frequencies, Raman Exp.: experimental Raman frequencies.}
\label{fig:6}%
\end{figure}

Utilizing THeSeuSS at various theoretical levels, including PBE+MBD-NL, DFTB3+MBD, and additional DFT functionals such as PBE, PBE0, PBE0+MBD-NL, B3LYP (as detailed in the Supporting Information), we successfully identified all 24 vibrational modes, ranging from strong and medium to weak intensity bands. However, it is notable that experimental IR and Raman spectra typically display only the stronger or medium intensity bands, often reporting fewer than 24 bands in the literature. Our simulated spectra, shown in Figure \ref{fig:6}, alongside previously established calculations and experimental frequencies, are presented.\cite{doi:10.1021/jp973397a, KUMAR20052741}

To ensure a quantitative comparison of our implementation, we calculated the MAPE and frequency shifts, both expressed as percentages, (see Tables \ref{table:2} and S2) relative to prior calculations rather than experimental data, which could lack complete spectral bands and potentially be misleading due to the presence of multiple conformers. This approach guarantees a precise, one-to-one comparison, crucial for an accurate assessment of our implementation. 

\begin{table}[h!]
\centering  
{\scriptsize
\begin{tabular}{|c c  c  | c c | c c|}
    \hline
    \multicolumn{3}{|c}{Frequencies (cm$^{-1}$)}& \multicolumn{2}{c}{Frequency shifts (\%)} & \multicolumn{2}{c|}{MAPE (\%)}\\ 
    \hline\hline
    PBE+MBD-NL & DFTB3+MBD & Ref.\cite{doi:10.1021/jp973397a} & PBE+MBD-NL & DFTB3+MBD & PBE+MBD-NL & DFTB3+MBD  \\  
    \hline
    \begin{tabular}{@{}c@{}} 87 \\ 161 \\ 272 \\ 455 \\ 499 \\ 607 \\ 643 \\ 808 \\ 822 \\ 976 \\ 1088 \\ 1106 \\ 1189 \\ 1206 \\ 1285 \\ 1388 \\ 1433 \\ 1585 \\ 1756 \\ 2889 \\ 2974 \\ 3417 \\ 3516 \\ 3633 \end{tabular} & \begin{tabular}{@{}c@{}} 113 \\ 241 \\ 371 \\ 624 \\ 674 \\ 849 \\ 901 \\ 1150 \\ 1166 \\ 1384 \\ 1516 \\ 1565 \\ 1673 \\ 1710 \\ 1721 \\ 1923 \\ 1982 \\ 2109 \\ 2406 \\ 3969 \\ 4069 \\ 4468 \\ 4669 \\ 5005  \end{tabular} & \begin{tabular}{@{}c@{}} 62 \\ 202 \\ 253 \\ 457 \\ 508 \\ 624 \\ 642 \\ 808 \\ 898 \\ 904 \\ 1113 \\ 1151 \\ 1166 \\ 1286 \\ 1359 \\ 1373 \\ 1425 \\ 1641 \\ 1787 \\ 2925 \\ 2962 \\ 3358 \\ 3429 \\ 3586 \end{tabular} & \begin{tabular}{@{}c@{}} 41.24 \\ 20.29 \\ 7.51 \\ 0.37 \\ 1.82 \\ 2.75 \\ 0.23 \\ 0 \\ 8.51 \\ 7.91 \\ 2.25 \\ 3.88 \\ 1.96 \\ 6.22 \\ 5.45 \\ 1.12 \\ 0.59 \\ 3.43 \\ 1.75 \\ 1.24 \\ 0.40 \\ 1.75 \\ 2.55 \\ 1.31 \end{tabular} & \begin{tabular}{@{}c@{}} 82.73 \\ 19.22 \\ 46.52 \\ 36.64 \\ 32.74 \\ 36.13 \\ 40.30 \\ 42.37 \\ 29.84 \\ 53.13 \\ 36.17 \\ 35.97 \\ 43.45 \\ 32.96 \\ 26.62 \\ 40.08 \\ 39.06 \\ 28.53 \\ 34.65 \\ 35.71 \\ 37.36 \\ 33.06 \\ 36.16 \\ 39.56 \end{tabular} & 5.19 & 38.29 \\
    \hline
\end{tabular}
}
\caption{Performance benchmark of THeSeuSS for conformer I of the glycine molecule in the gas phase, quantified by MAPE and frequency shifts expressed as percentages. This table lists frequencies calculated with THeSeuSS and compares them to those previously determined using DFT/B3LYP/aug-cc-pVDZ.\cite{doi:10.1021/jp973397a} Frequency shifts and MAPE are evaluated relative to these established reference data. Abbreviation: Ref.: reference data.}
\label{table:2} 
\end{table}

As demonstrated in Table \ref{table:2}, the simulated spectra at the PBE+MBD-NL level show good agreement with the reference data, particularly in accurately describing the vibrational frequencies. Some bands exhibit frequency shifts, likely due to the absence of anharmonic effects in our implementation. Similar frequency shifts are observed with hybrid functionals (PBE0, B3LYP as shown in Table S2), and MBD-NL does not affect the results, similar to the H$_{2}$O molecule in the gas phase case. Conversely, the DFTB3+MBD method reveals more significant deviations from the reference data, consistent with the expected limitations of the DFTB approach. Despite these deviations, we present DFTB data for glycine to provide a comprehensive analysis of our implementation.

It is noteworthy that the deviations between our simulated frequencies and the experimental data (Figure \ref{fig:6}) are consistent with those between the experimental data and the reference data used in our study of glycine, despite both sets of data originating from the same study.\cite{doi:10.1021/jp973397a} This alignment suggests that the discrepancies between our simulated frequencies and the experimental frequencies do not indicate deficiencies in our implementation. Instead, in the case of glycine molecule in gas phase, they reflect the inherent limitations of experimental techniques, which often fail to capture complete spectral bands and can be influenced by the presence of multiple conformers.

In addition to comparing our simulated results with reference data, a crucial aspect of our analysis is assigning specific frequencies to particular vibrations. For instance, the C=O stretching vibrations are identified at a calculated mode of approximately 1756 cm$^{-1}$. The mode at 1585 cm$^{-1}$ corresponds to O-H bending vibrations. The frequency range between 808 and 1400 cm$^{-1}$ encompasses C-O and C-C stretching vibrations, with a subset from 822 to 1189 cm$^{-1}$ also encompassing CNH bending vibrations. Additionally, N-H stretching vibrations appear between 3417 and 3516 cm$^{-1}$, and O-H stretching is observed at 3633 cm$^{-1}$. Low frequency ranges from 87 to 643 cm$^{-1}$ are associated with various torsional and bending vibrations, including NCC-O and NCC=O torsion, NCC bending, CCNH torsion, and CCOH torsion and bending. These assignments, based on our PBE+MBD-NL data, align with previously reported mode assignments.\cite{doi:10.1021/jp973397a, KUMAR20052741}

To evaluate and benchmark the performance of THeSeuSS on periodic systems, we chose phase I of solid ammonia, which crystallizes in the cubic P2$_{1}$3 structure, as a test case. Despite numerous reports over the past fifty years detailing the IR and Raman spectra of solid ammonia's different phases, these studies often present conflicting spectral characteristics.\cite{10.1063/1.1669380, 10.1063/1.453383, ZHENG2007229, doi:10.1366/0003702804731339, GAUTHIER1986218, BINBREK1972421, huang2020ab, https://doi.org/10.1002/jrs.710} Notably, to the best of our knowledge, there are no previously computed IR and Raman spectra for phase I of solid ammonia.

\begin{figure}[h!]%
\centering
\subfigure[IR spectra of ammonia in solid phase (phase I).]{
\includegraphics[height=2.49in]{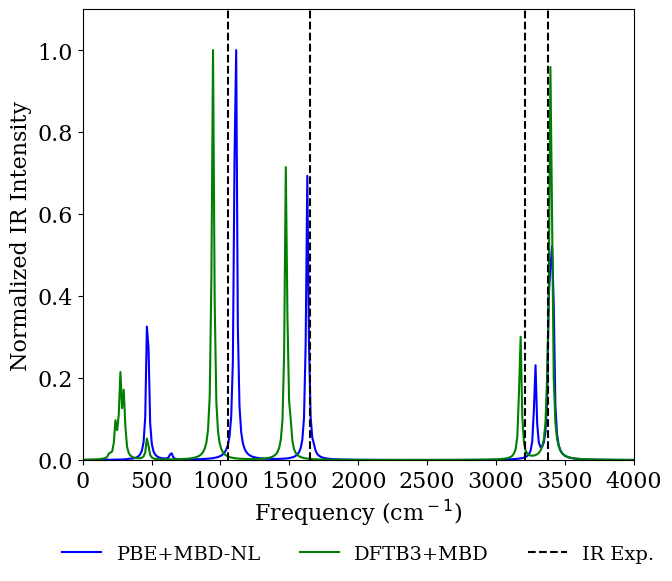}}%
\qquad
\subfigure[Raman spectra of ammonia in solid phase (phase I).]{
\includegraphics[height=2.49in]{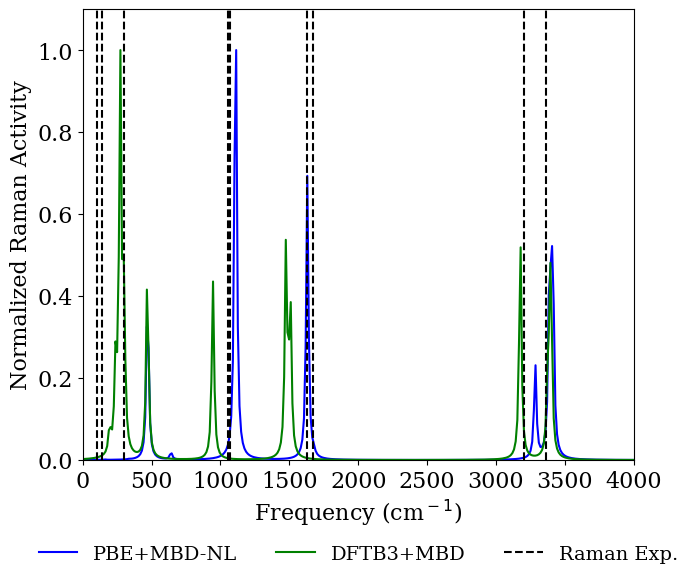}}%
\caption{Vibrational spectra of ammonia in solid phase (Phase I): (a) IR spectra calculated at both the PBE+MBD-NL and DFTB3+MBD levels, alongside experimental IR frequencies.\cite{10.1063/1.1669380} (b) Raman spectra similarly calculated and compared with experimental Raman frequencies.\cite{10.1063/1.453383} Abbreviations: IR Exp.: experimental IR frequencies, Raman Exp.: experimental Raman frequencies.}
\label{fig:7}%
\end{figure}

In assessing the accuracy of our implementation to describe vibrational spectra, we compare our simulated spectra with the most recent experimental data, as illustrated in Figure \ref{fig:7} and detailed in Table \ref{table:3}.\cite{10.1063/1.1669380, 10.1063/1.453383} It is important to note that these experimental results primarily focus on the most intense bands, omitting a comprehensive list of all vibrational modes. Consequently, we opted for a distinct approach to quantitative analysis, which differs from our methodology for analyzing the vibrational frequencies of H$_{2}$O and glycine molecules in the gas phase. Our simulations capture frequencies corresponding to both strong and weak bands. However, in Table \ref{table:3}, we specifically list the simulated frequencies of the most intense bands, with a few exceptions for weak bands in the Raman spectrum, enabling a one-to-one comparison with the experimental data.

\begin{table}[h!]
\centering  
{\scriptsize
\begin{tabular}{|c c  c  | c c | c c|}
    \hline
    \multicolumn{3}{|c}{Frequencies (cm$^{-1}$)}& \multicolumn{2}{c}{Frequency shifts (\%)} & \multicolumn{2}{c|}{MAPE (\%)}\\ 
    \hline\hline
    PBE+MBD-NL & DFTB3+MBD & Ref. & PBE+MBD-NL & DFTB3+MBD & PBE+MBD-NL & DFTB3+MBD  \\  
    \hline
    \multicolumn{7}{|c|}{IR} \\
    \hline
    \begin{tabular}{@{}c@{}} - \\ 466 ($\nu_\text{L}$) \\ 1111 ($\nu_2$) \\ 1631 ($\nu_4$) \\ 3286 ($\nu_1$) \\ 3390 ($\nu_3$) \end{tabular} & \begin{tabular}{@{}c@{}} 273 (R) \\ 470 ($\nu_\text{L}$) \\ 946 ($\nu_2$) \\ 1476 ($\nu_2$+$\nu_\text{L}$) \\ 3177 ($\nu_1$) \\ 3398 ($\nu_3$) \end{tabular} & \begin{tabular}{@{}c@{}} - \\ - \\ 1057 ($\nu_2$)\cite{10.1063/1.1669380} \\ 1650 ($\nu_4$)\cite{10.1063/1.1669380} \\ 3210 ($\nu_1$)\cite{10.1063/1.1669380} \\ 3375 ($\nu_3$)\cite{10.1063/1.1669380} \end{tabular} & \begin{tabular}{@{}c@{}} - \\ - \\ 5.11 \\ 1.15 \\ 2.37 \\ 0.44 \end{tabular} & \begin{tabular}{@{}c@{}} - \\ - \\ 10.50 \\ - \\ 1.03 \\ 0.68 \end{tabular} & 2.27 & 4.07 \\
    \hline
     \multicolumn{7}{|c|}{Raman} \\
    \hline
    \begin{tabular}{@{}c@{}} - \\ 130 (T/W) \\ 162 (T/W) \\ - \\ 471 ($\nu_\text{L}$) \\ 1111 ($\nu_2$) \\ 1142 ($\nu_2$+$\nu_\text{L}$) \\ 1631 ($\nu_4$) \\ 1677 ($\nu_4$) \\ 3286 ($\nu_1$) \\ 3389 ($\nu_3$) \\ 3412 ($\nu_1$+$\nu_\text{L}$)\end{tabular} & \begin{tabular}{@{}c@{}} - \\ - \\ 161 (T/W) \\ 273 (R) \\ 470 ($\nu_\text{L}$) \\ 946 ($\nu_2$) \\ 953 ($\nu_2$) \\ 1476 ($\nu_2$+$\nu_\text{L}$) \\ 1507 ($\nu_2$+$\nu_\text{L}$) \\ 3177 ($\nu_1$) \\ 3398 ($\nu_3$) \\ - \end{tabular} & \begin{tabular}{@{}c@{}} 107 (T)\cite{10.1063/1.453383} \\ 139 (T)\cite{10.1063/1.453383} \\ 160 (T)\cite{BINBREK1972421} \\ 297 (L)\cite{10.1063/1.453383} \\ - \\ 1055 ($\nu_2$)\cite{10.1063/1.453383} \\ 1070 ($\nu_2$)\cite{10.1063/1.453383} \\ 1630 ($\nu_4$)\cite{10.1063/1.453383} \\ 1675 ($\nu_4$)\cite{10.1063/1.453383} \\ 3200 ($\nu_1$)\cite{10.1063/1.453383} \\ 3366 ($\nu_3$)\cite{10.1063/1.453383} \\ - \\ \end{tabular} & \begin{tabular}{@{}c@{}} - \\ 6.47 \\ 1.25 \\ - \\ - \\ 5.31 \\ 6.73 \\ 0.06 \\ 0.12 \\ 2.69 \\ 0.68 \\ - \end{tabular} & \begin{tabular}{@{}c@{}} - \\ - \\ 0.62 \\ 8.08 \\ - \\ 10.33 \\ 10.93 \\ 9.45 \\ 10.03 \\ 0.72 \\ 0.95 \\ - \end{tabular} & 2.91 & 6.39 \\
    \hline
\end{tabular}
}
\caption{Performance benchmark of THeSeuSS for phase I of solid ammonia, quantified by MAPE and frequency shifts expressed as percentages. This table displays frequencies calculated with THeSeuSS, comparing them against experimental data.\cite{10.1063/1.1669380, 10.1063/1.453383} Frequency shifts and MAPE are calculated relative to this experimental data. Abbreviations: Ref.: reference data, R: rotational mode, T: translational mode, L: librational mode, $\nu_\text{L}$: lattice mode, $\nu_1$: symmetric stretching, $\nu_2$: symmetric bending, $\nu_3$: antisymmetric stretching, $\nu_4$: antisymmetric bending, W: weak band.}
\label{table:3} 
\end{table}

More specifically, in the case of IR spectra analyzed using PBE+MBD-NL, the primary modes show close alignment between the computed spectra from THeSeuSS and the experimental spectra. Notably, there is an additional main mode present in the computed spectra that is absent in the experimental data. However, the frequency shifts and MAPE for the four modes that can be compared across both datasets demonstrate good agreement. Furthermore, as indicated in Table \ref{table:3}, the assignments of these modes correspond well between the computed and experimental spectra.

Conversely, when using DFTB3+MBD, the mode assignments do not align as closely, and the frequency shift for the 946 cm$^{-1}$ mode is significant. Additionally, two more main modes are observed at lower frequencies. It is crucial to note that a low MAPE value, calculated using only the three modes that correspond between the experimental and simulated spectra, does not necessarily reflect accurate spectrum description. These deviations are consistent with the known limitations of the DFTB approach, as previously discussed. Nevertheless, DFTB3+MBD demonstrates improved performance in analyzing the IR spectrum of solid ammonia compared to its performance with H$_{2}$O and glycine molecules in the gas phase.

Similarly, in the case of Raman spectra, PBE+MBD-NL demonstrates superior performance, particularly evident in its treatment of high-frequency modes, where it demonstrates good agreement with experimental data in terms of frequency shifts. Although our simulated spectra exhibit weak bands in the low-frequency modes, we include them in our analysis for a more comprehensive understanding. Notably, the mode at 160 cm$^{-1}$, though not prevalent in recent experimental spectra, has been reported by Binbrek \textit{et al.}\cite{BINBREK1972421} We speculate that its absence in recent publications could be attributed to its association with a weak band. It is conceivable that employing the PBE0+MBD-NL method could further enhance the agreement of these low-frequency modes.

To further examine THeSeuSS's efficiency with more complex systems, we calculated the IR spectrum of solid ibuprofen, a widely used medication for managing and treating inflammatory diseases. Solid ibuprofen crystallizes in the P 2$_{1}$/c space group. While numerous reports exist on the experimental and theoretical vibrational spectra of the molecular form of ibuprofen, detailed studies on the spectra of solid ibuprofen are scarce. In particular, to the best of our knowledge, no comprehensive study includes a detailed report of the Raman vibrational spectra and the assignments of vibrational modes for solid ibuprofen. Therefore, we only present the IR spectrum of solid ibuprofen, as it allows for comparison with previously measured experimental IR spectra and identified experimental vibrational frequencies.\cite{https://doi.org/10.1002/crat.201100394}

\begin{figure}[h!]%
\centering
\includegraphics[height=4.0in]{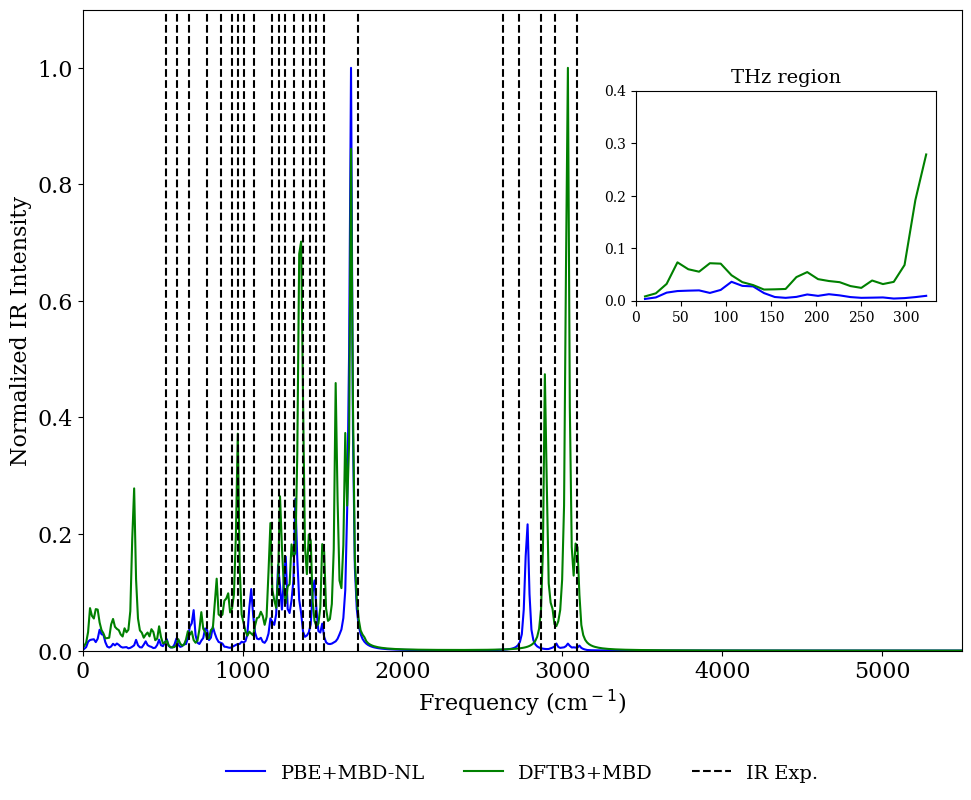}%
\caption{IR spectrum of ibuprofen in solid phase calculated at both the PBE+MBD-NL and DFTB3+MBD levels, alongside experimental IR frequencies.\cite{https://doi.org/10.1002/crat.201100394} Abbreviations: IR Exp.: experimental IR frequencies.}
\label{fig:8}%
\end{figure}

In Figure \ref{fig:8} we present the simulated IR spectra of solid ibuprofen alongside the experimental IR frequencies. The displayed vibrational frequencies correspond to medium, strong and very strong bands, while frequencies associated with weak and very weak bands are omitted. This Figure demonstrates that THeSeuSS effectively reproduces the IR spectrum of solid ibuprofen, a conclusion further supported by the frequency shifts expressed as percentages shown in Table \ref{table:4}. This indicates that PBE+MBD-NL accurately calculates the IR frequencies of the system. Notably, the simulated vibrational frequencies calculated at the DFTB3+MBD level also show better agreement with the experimental data compared to the previous systems we examined. The reasons behind this outcome warrant further investigation, as we hypothesize that the crystal structure of ibuprofen and the resulting molecular interactions may play a significant role. Future work will aim to elucidate the underlying factors contributing to this phenomenon. Furthermore, the assignment of modes corresponds well between experimental and computed spectra. The excellent comparison of our simulated spectra with the experimental counterparts showcases the efficiency of TheSeuSS in simulating vibrational spectra, even for complex systems like ibuprofen. While this study provides a comprehensive analysis of the high frequencies region, the THz region (shown in the inset of Figure \ref{fig:8}) warrants further investigation. Future work will explore this aspect in greater detail.

\begin{table}[h!]
\centering  
{\scriptsize
\begin{tabular}{|c c  c c | c c |}
    \hline
    \multicolumn{3}{|c}{Frequencies (cm$^{-1}$)}& Assignments & \multicolumn{2}{c|}{Frequency shifts (\%)} \\ 
    \hline\hline
    PBE+MBD-NL & DFTB3+MBD & Ref.\cite{https://doi.org/10.1002/crat.201100394} &  & PBE+MBD-NL & DFTB3+MBD \\  
    \hline
    \begin{tabular}{@{}c@{}} 521 \\ 586 \\ 691 \\ 774 \\ 814 \\ 935 \\ 966 \\ 1010 \\ 1054 \\ 1126 \\ 1174 \\ 1226 \\ 1270 \\ 1330 \\ - \\ 1450 \\ - \\ 1498 \\ - \\ - \\ 1679 \\ - \\ 2661 \\ 2783 \\ - \\ 2963 \\ 3087 \\ 3107 \end{tabular} & \begin{tabular}{@{}c@{}} 526 \\ 598 \\ 658 \\ 742 \\ 838 \\ 910 \\ 970 \\ - \\ 1042 \\ 1114 \\ 1174 \\ 1234 \\ - \\ 1306 \\ 1366 \\ 1426 \\ - \\ 1498 \\ 1583 \\ 1643 \\ 1679 \\ - \\ - \\ - \\ 2891 \\ 3035 \\ 3095 \\ - \end{tabular} & \begin{tabular}{@{}c@{}} 522 \\ 588 \\ 668 \\ 779 \\ 866 \\ 936 \\ 970 \\ 1008 \\ 1071 \\ - \\ 1183 \\ 1231 \\ 1268 \\ 1321 \\ 1380 \\ 1420 \\ 1462 \\ 1507 \\ - \\ - \\ - \\ 1721 \\ 2632 \\ 2728 \\ 2869 \\ 2955 \\ 3090 \\ - \end{tabular} & \begin{tabular}{@{}c@{}} CH$_{2}$ in plane rocking \\ C...C def. \\ C-H out of plane def. \\ CH$_{2}$ rocking \\ C-H out of plane vib. \\ CH$_{3}$ rocking vib. \\ C-O-C str. \\ C-H in plane def. \\ =C-H in plane def. \\ =C-H in plane def. \\ C-O str. \\ C...C str. \\ =C-H in plane def. \\ OH in plane def. \\ CH$_{3}$ sym. str. \\ CH-CO def. \\ CH$_{3}$ antisym. def., CH$_{2}$ scissoring \\ aromatic C=C str. \\ aromatic C=C str. \\ aromatic C=C str. \\ C=O str. \\ C=O str. \\ O-H...O valance str. combination \\ O-H...O valance str. combination \\ CH$_{2}$ sym. str. \\ CH$_{3}$ antisym. str. \\ CH$_{2}$ antisym. str. \\ CH$_{2}$ antisym. str. \end{tabular} & \begin{tabular}{@{}c@{}} 0.19 \\ 0.34 \\ 3.44 \\ 0.64 \\ 6.00 \\ 0.11 \\ 0.41 \\ 0.20 \\ 1.59 \\ - \\ 0.76 \\ 0.41 \\ 0.16 \\ 0.68 \\ - \\ 2.11 \\ - \\ 2.46 \\ - \\ - \\ - \\ - \\ 1.10 \\ 2.02 \\ - \\ 0.27 \\ 0.10 \\ - \end{tabular} & \begin{tabular}{@{}c@{}} 0.77 \\ 1.70 \\ 1.50 \\ 4.75 \\ 3.23 \\ 2.78 \\ 0 \\ - \\ 2.71 \\ - \\ 0.76 \\ 0.24 \\ - \\ 1.14 \\ 1.01 \\ 0.42 \\ - \\ 0.60 \\ - \\ - \\ - \\ - \\ - \\ - \\ 0.77 \\ 2.71 \\ 0.16 \\ - \end{tabular} \\
    \hline
     & & & & \multicolumn{2}{|c|}{MAPE (\%)}\\
     \hline
     & & & & 1.21 & 1.48 \\
     \hline
    \end{tabular}
}
\caption{Performance benchmark of THeSeuSS of solid ibuprofen, quantified by MAPE and frequency shifts expressed as percentages. This table displays IR frequencies calculated with THeSeuSS, comparing them against experimental data.\cite{https://doi.org/10.1002/crat.201100394} Frequency shifts and MAPE are calculated relative to this experimental data. Abbreviations: Ref.: reference data, def.: deformation, vib.: vibration, str.: stretching, sym.: symmetric, antisym.: antisymmetric.}
\label{table:4} 
\end{table}

It is crucial to note that we have not yet discussed spectral intensities nor compared our simulated results with experimental intensities. Discrepancies between simulated and experimental intensities are expected, primarily due to variations in the settings of experimental apparatuses. Specifically, in Raman spectroscopy, the accurate calculation of Raman activity using Equation \ref{eq.15} requires that the incident beam's direction, polarization orientation, and observation direction be mutually perpendicular. However, this ideal alignment is not always achievable in experimental setups, leading to potential disparities between experimental and calculated intensities.

Consequently, IR intensity and Raman activity should be regarded as theoretical estimates. Discrepancies from experimental intensities are anticipated and should not be seen as limitations of our implementation. It is worth mentioning that this approach is common across most computational codes that calculate IR and Raman intensities, reflecting a standard practice in theoretical spectroscopy.

\subsection{Conclusions}

We have developed THeSeuSS, an automated computational platform designed to improve the efficiency of simulating IR and Raman spectra. This Python-based package adeptly captures the complex details of vibrational spectra, requiring only the system's crystal structure and minimal additional input from the user. THeSeuSS is capable of generating spectra using both DFT and DFTB methods. It employs a static approach to calculate vibrational frequencies and leverages the calculated dipole moments, cartesian polarization, and polarizability matrices to determine IR intensity and Raman activity, delivering high-fidelity spectral data for both periodic and non-periodic systems.

THeSeuSS interfaces with FHI-aims for DFT-level calculations and with DFTB+ for DFTB-level calculations, ensuring broad compatibility. For phonon frequency analysis in periodic systems, it interfaces with PHONOPY, whereas for non-periodic systems, it utilizes specially developed built-in functions. THeSeuSS offers a comprehensive suite of functionalities: (1) conversion of experimental geometry inputs, (2) determination of space group for both experimental and optimized structures, (3) geometry optimization, (4) vibrational or phonon frequency calculations, (5) calculation of properties such as dipole moments and polarizability matrices, (6) calculation of IR intensity and Raman activity, and (7) spectra generation. Users can operate these components either simultaneously or individually, tailoring the process to their specific needs. Additionally, THeSeuSS can also compute THz spectra, focusing solely on the calculation of spectral frequencies.

To minimize the computational time, we have implemented parallelization across several key components of the code. Specifically, the single point calculations required for the finite difference method, used in determining vibrational/phonon frequencies and other properties, are fully parallelized. Additionally, by integrating PHONOPY and leveraging its ability to recognize the symmetry of the crystal structure, we have introduced an efficiency enhancement. This integration reduces the number of necessary single-point calculations by limiting them to only those atomic coordinates within the irreducible representation.

Our evaluations across diverse molecular and solid systems; including non-periodic systems like H$_{2}$O and glycine molecules, as well as the periodic systems phase I of solid ammonia, and solid ibuprofen highlight THeSeuSS's ability to reliably reproduce IR and Raman spectra. These results closely align with both experimental data and previously established calculations.

Our developed code, THeSeuSS, has implications across multiple industries that depend on vibrational spectroscopy, including pharmaceuticals and materials science. These fields require precise molecular characterization, while streamlining the spectrum calculation process and offering capabilities to handle a variety of systems.

As we move forward, we anticipate further enhancements to THeSeuSS, integrating additional computational methods and expanding its capabilities through machine learning techniques. These improvements aim to expedite the simulation of vibrational spectra and enable the handling of larger, more complex systems, such as biological molecules and polymers. Ongoing development, driven by user feedback and technological advancements will ensure that THeSeuSS remains at the forefront of computational spectroscopy tools.

\subsection{Computational details}

PBE, PBE+MBD-NL, PBE0, PBE0+MBD-NL and B3LYP calculations were performed using the all-electron numeric-atom-centered orbital code FHI-aims (Fritz Haber Institute ab initio molecular simulations).\cite{PhysRevLett.77.3865, PhysRevLett.124.146401, blum2009ab, knuth2015all, ren2012resolution, YU2018267, HAVU20098367, ihrig2015accurate} Scalar relativistic effects were included using the zero-order regular approximation (ZORA), in the case of PBE and PBE+MBD-NL. The tight species default settings in FHI-aims were applied for all numerical atom-centered basis functions and integration grids. Convergence criteria of 10$^{-6}$ eV for total energy, 10$^{-7}$ electrons/Å$^{3}$ for charge density, 10$^{-5}$ eV/Å for the sum of eigenvalues, and 10$^{-4}$ eV for forces were employed. In cases of cell and geometry optimization, the maximum residual force component per atom, below which the geometry relaxation is considered converged was set to 10$^{-5}$ eV/Å, while the maximum acceptable energy increase per relaxation step was also set to 10$^{-5}$ eV. These values were determined after performing convergence tests for total energy, pressure, stresses, and atomic forces. For ammonia and ibuprofen, the Brillouin zone was sampled with a 4$\times$4$\times$4 and 2$\times$3$\times$3 Monkhorst-Pack k-points grid, respectively.\cite{monkhorst1976special}

DFTB3+MBD calculations were performed using the DFTB+ simulation package.\cite{10.1063/1.5143190, PhysRevLett.108.236402} The 3ob-3-1 parameter set, part of the third-order parametrization for organic and biological systems (3OB), was employed.\cite{doi:10.1021/ct300849w, doi:10.1021/ct401002w, doi:10.1021/jp506557r, doi:10.1021/ct5009137} The self-consistent cycle stopping criterion was set at 10$^{-7}$ eV. During cell and geometry optimization, the process is halted when the maximal absolute value of the force component drops below 10$^{-5}$ eV/Å. For ammonia and ibuprofen, the Brillouin zone was sampled using a 5$\times$5$\times$5 and 2$\times$3$\times$3 Monkhorst-Pack k-points grid, respectively.\cite{monkhorst1976special} MBD interactions were evaluated using a simpler 1$\times$1$\times$1 k-point mesh with $\beta=0.83$.

In both periodic and non-periodic cases, the displacement values used in the finite displacement method adhere to PHONOPY's default settings: 0.01 Å for simulations run with FHI-aims and 0.005 Å for those utilizing DFTB+. For the graphical representation of the vibrational spectra, we employed Lorentzian broadening with the full width at half maximum (FWHM) set to 10.0.

\section{Supporting Information}

Additional benchmark data for the performance of THeSeuSS, including computed frequencies, frequency shifts expressed as percentages, and MAPE for the H$_{2}$O and glycine molecules in the gas phase.

\begin{acknowledgement}

The authors extend their gratitude to Dr. Olympia Giannou for her insightful discussions regarding coding queries. Special thanks are also due to Dr. Matteo Barborini for his valuable insights into the code and to Dr. Georgios Kafanas for his guidance on parallelization techniques using slurm. The authors are also grateful to Kyunghoon Han for his assistance with mathematical formulations. Lastly, heartfelt thanks are extended to Dr. Benjamin Hourahine and Dr. Mariana Rossi for their invaluable discussions on the intricacies of the DFTB+ and FHI-aims codes, respectively. The calculations presented in this paper were carried out using the HPC facilities of the University of Luxembourg (see hpc.uni.lu),\cite{varrette2014management} and the Luxembourg national supercomputer MeluXina. A.B. and A.T. gratefully acknowledge the HPC and LuxProvide teams for their expert support. A.T. acknowledges funding from Janssen Pharmaceuticals under the project THESIS3, the FNR BroadApp grant and MACHINE-DRUG grant.

\end{acknowledgement}


\bibliography{achemso-demo}

\end{document}